\documentstyle[12pt,a41,epsfig]{article}
\newcommand{\bq}{\begin{equation}}
\newcommand{\eq}{\end{equation}}

\newcommand\ka{\kappa_1}
\newcommand\kb{\kappa_2}
\newcommand\kap{\kappa_1'}
\newcommand\kbp{\kappa_2'}
\newcommand\xx{\tilde{x}}
\newcommand\AAA{\alpha_1}
\newcommand\AB{\alpha_2}
\newcommand\Kvec{\mbox{\boldmath $K$}}
\newcommand\Pvec{\mbox{\boldmath $P$}}

\newcommand\Vvec{\mbox{\boldmath $V$}}
\newcommand\bx{\overline{x}}
\newcommand\by{\overline{y}}
\newcommand\FFA{\mbox{$\widetilde{F}^a$}}

\begin{document}
\sloppy
\thispagestyle{empty}
\begin{flushleft}
DESY 97--074\\
NTZ  14/97  \\
{\tt hep-ph/9705264}\\
April 1997 \\
Phys. Lett. {\bf B} in print
\end{flushleft}

\mbox{}
\vspace*{\fill}
\begin{center}
{\LARGE\bf On the Evolution Kernels of Twist 2 Light-Ray}

\vspace{2mm}
{\LARGE\bf Operators
for Unpolarized and Polarized }\\

\vspace{2mm}
{\LARGE\bf Deep Inelastic Scattering}
\\

\vspace{2em}
\large
Johannes Bl\"umlein$^a$, Bodo Geyer$^b$, and Dieter Robaschik$^a$,
\\
\vspace{2em}
{\it $^a$~DESY -- Zeuthen,}
 \\
{\it Platanenallee 6,
D--15735 Zeuthen, Germany}\\

\vspace{2em}
{\it $^b$~Naturwissenschaftlich--Theoretisches Zentrum der
Universit\"at Leipzig,}\\
{\it Augustusplatz 10, D--04109 Leipzig, Germany}
 \\
\end{center}
\vspace*{\fill}
\begin{abstract}
\noindent
The non--singlet and singlet evolution kernels of  the twist--2
light--ray operators for unpolarized and polarized deep inelastic
scattering are calculated in ${\cal O}(\alpha_s)$ for the general case of
virtualities $q^2, q'^2 \neq 0$. Special  cases as the kernels
for the general single--variable evolution equation and the
Altarelli--Parisi and Brodsky--Lepage limits are derived from these
results.
\end{abstract}
\vspace*{\fill}
\newpage
\noindent
\section{Introduction}
The study of the Compton amplitude for scattering a virtual photon
off a hadron is one of the basic tools in QCD to understand the
short--distance behavior of the theory.
The Compton amplitude for the general
case of  non--forward scattering is given by
\begin{equation}
\label{COMP}
T_{\mu\nu}(p_+,p_-,Q) = i \int d^4x e^{iqx}
\langle p_2|T (J_{\mu}(x/2) J_{\nu}(-x/2))|p_1\rangle,
\end{equation}
where $p_+ = p_2 + p_1, p_- = p_2 - p_1 = q_1 - q_2, Q = (q_1 + q_2)/2,
p_1 + q_1 = p_2 + q_2$. The kinematics of the basic
process is depicted schematically in figure~1.
\begin{center}

\vspace{2cm}
\begin{picture}(120,20)(100,100) \centering
\put(0,-30){\epsfig{file=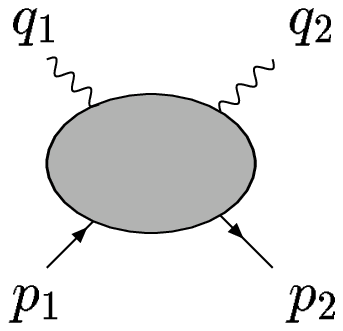,width=16cm}}
\end{picture}
\noindent
\small
\end{center}
\begin{center}
{\sf Figure~1:~Kinematic notation for the Compton amplitude}
\normalsize
\end{center}
The time--ordered product  in eq.~(\ref{COMP}) can be
represented in terms of the operator product expansion. Here we use
the representation derived in ref.~\cite{AS,LEIP},
\begin{eqnarray}
\label{tpro}
T (J_{\mu}(x/2) J_{\nu}(-x/2)) &\approx&
\int_{-\infty}^{+\infty} d \kappa_-
\int_{-\infty}^{+\infty} d \kappa_+
\left [ C_a(x^2, \kappa_-, \kappa_ +, \mu^2) {S_{\mu\nu}}^{\rho\sigma}
\xx_{\rho} O_{\sigma}^a (\kappa_+ \xx, \kappa_- \xx, \mu^2) \right.
\nonumber\\
& & \left.
~~~~~~~~~~~~~~~~~~~~+
C_{a,5}(x^2, \kappa_-, \kappa_ +, \mu^2)
{\varepsilon_{\mu\nu}}^{\rho\sigma}
\xx_{\rho} O_{5, \sigma}^a (\kappa_+ \xx, \kappa_- \xx, \mu^2) \right ],
\end{eqnarray}
with $S_{\mu\nu\rho\sigma} = g_{\mu\nu} g_{\rho\sigma}
- g_{\mu\rho} g_{\nu\sigma} - g_{\mu\sigma} g_{\nu\rho}$ and
$\varepsilon_{\mu\nu\rho\sigma}$ denoting the Levi--Civita symbol.
The light--like vector
\begin{equation}
\xx = x + r (x.r/r.r) \left [ \sqrt{1 - x.x r.r / (x.r)^2} - 1 \right ]
\end{equation}
is related to $x$ and a subsidiary four--vector
$r$.  $C_{a (5)}$ denote the respective coefficient functions.
Unlike the local operator product expansion~\cite{OPEL}, which is mainly
used
in the case of forward scattering, the representation
eq.~(\ref{tpro})
allows to derive the anomalous dimensions and coefficient functions
also for the non--forward case for general values of the virtuality
$q_2^2$.

In this paper we present the results of a calculation of the non--singlet
and singlet twist~2~\footnote{The notion of twist is not quite
unambiguous in this context. It is rather used to label the type of
operators
being considered, for which in the limit $p_2 \rightarrow p_1$
the notion applies~\cite{GROSS}.}
anomalous dimensions for this general case both for
unpolarized and polarized scattering. Details of the calculation will
be published elsewhere~\cite{BGR1}.
We derive general evolution equations both for the operators and
matrix elements in the non--forward case.
Furthermore we derive
a series of special cases discussed in the literature previously and
compare the results.
\section{The Evolution Kernels}
We consider the flavor non--singlet and singlet operators
in the axial gauge $\tilde{x}_{\mu} A^{\mu} = 0$
\begin{eqnarray}
O^{\rm NS}(\ka,\kb) &=&
\frac{i}{2}
\left [
\overline{\psi_a}(\ka\xx)
\lambda_f
\gamma_{\mu} \xx^{\mu} \psi_a(\kb \xx) -
\overline{\psi_a}(\kb\xx)
\lambda_f
\gamma_{\mu} \xx^{\mu} \psi_a(\ka \xx) \right ]\\
O^{\rm NS}_5(\ka,\kb) &=&
\frac{i}{2}
\left [
\overline{\psi_a}(\ka\xx) \gamma_5
\lambda_f
\gamma_{\mu} \xx^{\mu} \psi_a(\kb \xx) +
\overline{\psi_a}(\kb\xx) \gamma_5
\lambda_f
\gamma_{\mu} \xx^{\mu} \psi_a(\ka \xx) \right ]\\
O^q_{\rm     }(\ka,\kb) &=&
\frac{i}{2} \left [
\overline{\psi_a}(\ka\xx)
\gamma_{\mu} \xx^{\mu} \psi_a(\kb \xx) -
\overline{\psi_a}(\kb\xx)
\gamma_{\mu} \xx^{\mu} \psi_a(\ka \xx) \right ]\\
O^q_{\rm 5   }(\ka,\kb) &=&
\frac{i}{2}
\left [
\overline{\psi_a}(\ka\xx) \gamma_5
\gamma_{\mu} \xx^{\mu} \psi_a(\kb \xx) +
\overline{\psi_a}(\kb\xx) \gamma_5
\gamma_{\mu} \xx^{\mu} \psi_a(\ka \xx) \right ]\\
O^G_{\rm     }(\ka,\kb) &=&
\xx^{\mu} {F_{a\mu}}^{\nu}(\ka\xx) \xx^{\mu'} {F^a}_{\mu'\nu}(\kb\xx)\\
O^G_{\rm 5   }(\ka,\kb) &=& \frac{1}{2} \left [
\xx^{\mu} {F_{a \mu}}^{\nu}(\ka\xx) \xx^{\mu'}
\FFA_{\mu'\nu}(\kb\xx)
- \xx^{\mu} {F^a}_{\mu\nu}(\kb\xx) \xx^{\mu'}
{\FFA_{\mu'}}{\vspace*{-1.2mm}^{\nu}}(\ka\xx)
\right ],
\end{eqnarray}
where $\psi_a$ denotes the quark field-
and $F_{\mu\nu}^a$ the gluon field strength operators, respectively.
The dual operator to $F_{\mu\nu}^a$ is
\begin{equation}
\widetilde{F}_{\mu\nu} = \frac{1}{2}
\varepsilon_{\mu\nu\rho\sigma} F^{\rho\sigma},
\end{equation}
\begin{equation}
\kappa_1 = \kappa_+ - \kappa_-~~~~~~~~~\kappa_2 = \kappa_+ + \kappa_-,
\end{equation}
and
$\lambda_f$ is a generator of the flavor group $SU(N_f)$, $N_f$  being
the number of active quark flavors.

The renormalization group equation implies
the following evolution equations for these operators~:
\begin{eqnarray}
\label{evo1}
\mu^2 \frac{d}{d \mu^2} O^{\rm NS}_{(5)}(\ka,\kb) &=&
\frac{\alpha_s(\mu^2)}{2 \pi}
\int_0^1 d \AAA \int_0^1 d \AB \theta(1 - \AAA - \AB)
K^{\rm NS}(\AAA,\AB)
O^{\rm NS}_{(5)}(\kap,\kbp), \\
\mu^2 \frac{d}{d \mu^2} \left (\begin{array}{c}
O^q(\ka,\kb) \\
\label{evo2}
O^G(\ka,\kb) \end{array} \right ) &=&
\frac{\alpha_s(\mu^2)}{2 \pi}
\int_0^1 d \AAA \int_0^1 d \AB \theta(1 - \AAA - \AB)
\Kvec(\AAA,\AB)
\left (\begin{array}{c}
O^q(\kap,\kbp) \\
O^G(\kap,\kbp) \end{array} \right ),
\nonumber \\
\end{eqnarray}
with $\alpha_s = g_s^2/(4\pi)$ the strong coupling constant, $\mu$
the renormalization scale, and  $\kap = \ka(1-\AAA)+\kb\AAA,
\kbp = \kb(1-\AB)+\ka\AB$.  $\Kvec$
denotes the
matrix of the singlet
evolution kernels in  the unpolarized case. The singlet evolution
equations for the polarized case are obtained replacing $O^{q,G}$ by
$O^{q,G}_{5}$ and $\Kvec$ by $\Delta \Kvec$. As far as relations are
concerned which are valid both for the unpolarized and polarized case
under this replacement, we will, for brevity, only give that for the
unpolarized case  in the following. The matrices of the
singlet kernels are
\begin{equation}
\Kvec = \left ( \begin{array}{ll} K^{qq} & K^{qG} \\
                                  K^{Gq} & K^{GG} \end{array} \right )
~~~~{\rm and}~~~~\Delta \Kvec = \left ( \begin{array}{ll} \Delta K^{qq} &
\Delta K^{qG} \\ \Delta K^{Gq} & \Delta K^{GG} \end{array} \right )~,
\end{equation}
respectively.
The non--singlet kernels obey
$K^{\rm NS} = \Delta K^{\rm NS} = K^{qq} = \Delta K^{qq}$.

In the unpolarized case the kernels are given by
\begin{eqnarray}
\label{eqK1}
K^{qq}(\AAA,\AB) &=&  C_F \left \{ 1 - \delta(\AAA) - \delta(\AB)
+ \delta(\AAA) \left [ \frac{1}{\AB}\right]_+
+ \delta(\AB) \left [ \frac{1}{\AAA}\right]_+
+ \frac{3}{2} \delta(\AAA)\delta(\AB) \right \}\\
K^{qG}(\AAA,\AB) &=&  - 2 N_f T_R \kappa_-
\left \{ 1 - \AAA - \AB + 4 \AAA\AB
\right \} \\
K^{Gq}(\AAA,\AB) &=&  - C_F \frac{1}{\kappa_- - i \varepsilon}
\left \{  \delta(\AAA) \delta(\AB) + 2 \right \} \\
\label{eqK4}
K^{GG}(\AAA,\AB) &=&  C_A \left \{ 4 ( 1 - \AAA -\AB) + 12 \AAA \AB
\right.
\\  & &~~+ \left.
\delta(\AAA) \left (
 \left [\frac{1}{\AB} \right ]_+         - 2
+ \AB \right )
+ \delta(\AB) \left ( \left [\frac{1}{\AAA} \right ]_ +        - 2
+ \AAA \right )  \right \}
+ \frac{\beta_0}{2} \delta(\AAA)\delta(\AB),   \nonumber
\end{eqnarray}
in ${\cal O}(\alpha_s)$
where $C_F = (N_c^2-1)/2N_c \equiv 4/3, T_R = 1/2, C_A = N_c \equiv 3$,
$\beta_0 = (11 C_A - 4 T_R N_f)/3$.
The $+$-prescription is defined as
\begin{equation}
\int_0^1 dx
 \left [f(x,y)\right]_+ \varphi(x) = \int_0^1 dx
f(x,y)
\left [\varphi(x) - \varphi(y) \right]~,
\end{equation}
where the singularity of $f$ is of the type $\sim 1/(x - y)$.

The  kernels for the polarized case are
\begin{eqnarray}
\label{eqDK1}
\Delta K^{qq}(\AAA,\AB) &=& K^{qq}(\AAA,\AB) \\
\Delta K^{qG}(\AAA,\AB) &=&  - 2 N_f T_R \kappa_-
\left \{ 1 - \AAA - \AB \right \} \\
\Delta K^{Gq}(\AAA,\AB) &=&  -  C_F \frac{1}{\kappa_- - i \varepsilon}
\left \{\delta(\AAA) \delta(\AB) - 2 \right \} \\
\label{eqDK4}
\Delta K^{GG}(\AAA,\AB) &=&  K^{GG}(\AAA,\AB) - 12 C_A   \AAA \AB~.
\end{eqnarray}
Whereas the kernels for the polarized case are derived for the first
time, those for the unpolarized case were found several years ago
already in refs.~\cite{BGR,BB}.
The kernels $K^{ij}$ and $\Delta K^{ij}$ determine the respective
evolutions of the operators
$O^{\rm NS}_{(5)}, O^{q}_{(5)}$, and $O^{G}_{(5)}$ in $O(\alpha_s)$.

The evolution equations may also be written for the corresponding
expectation values $\langle p_2|O^{q,G}_{(5)}|p_1\rangle$ to which
two--variable  quark and gluon partition functions are
associated by
\begin{eqnarray}
\frac{\langle p_1|O^{q}|p_2\rangle}{(i\xx p_+)} &=&
e^{-i\kappa_+ \xx p_-}
\int_{-\infty}^{+\infty} d z_+
\int_{-\infty}^{+\infty} d z_-
e^{-i\kappa_-( \xx p_+ z_+ + \xx p_- z_-)} F_{q}(z_-,z_+)\\
\frac{\langle p_1|O^{G}|p_2\rangle}{(i\xx p_+)^2} &=&
e^{-i\kappa_+ \xx p_-}
\int_{-\infty}^{+\infty} d z_+
\int_{-\infty}^{+\infty} d z_-
e^{-i\kappa_-( \xx p_+ z_+ + \xx p_- z_-)} F_{G}(z_-,z_+)~.
\end{eqnarray}
The evolution equations for the partition functions read
\begin{eqnarray}
\label{evo3}
\mu^2 \frac{d}{d \mu^2} F^{\rm NS}(z_+,z_-) &=&
\frac{\alpha_s(\mu^2)}{2 \pi}
\int_{-\infty}^{+\infty} \frac{d z'_+}{|z'_+|}
\int_{-\infty}^{+\infty} d z'_-
\widetilde{K}^{\rm NS}(\alpha_1,\alpha_2)
F^{\rm NS}(z'_+,z'_-)
\nonumber\\ \\
\mu^2 \frac{d}{d \mu^2}
\left (\begin{array}{c}
F^{q}(z_+,z_-) \\
F^{G}(z_+,z_-) \end{array}\right )
&=&
\frac{\alpha_s(\mu^2)}{2 \pi}
\int_{-\infty}^{+\infty} \frac{d z'_+}{|z'_+|}
\int_{-\infty}^{+\infty} d z'_-
\widetilde{\Kvec}(z_+,z_-;z'_+,z'_-)
\left (\begin{array}{c}
F^{q}(z'_+,z'_-) \\
F^{G}(z'_+,z'_-) \end{array}\right ), \nonumber \\
\end{eqnarray}
where  $F^{\rm NS}(z_+,z_-) = F^{q_i}(z_+,z_-) -
F^{\overline{q}_j}(z_+,z_-)$, and
\begin{equation}
\widetilde{K}^{ij}(\alpha_1,  \alpha_2)
 = \frac{1}{2} \int_0^1 dz''_+ \widetilde{O}^{ij}(z_+,z''_+)
\theta(1 - \alpha'_+) \theta(\alpha'_+) \theta(\alpha'_+  +  \alpha'_-)
   \theta(\alpha'_+  -  \alpha'_-) K^{ij}(\alpha'_1,  \alpha'_2)~,
\end{equation}
with $\widetilde{K}^{\rm NS} = \widetilde{K}^{qq}$ and
$\alpha'_{\rho} =  \alpha_{\rho}(z_+ \rightarrow z''_+)$,
\begin{equation}
\alpha_1 = \frac{\alpha_+ + \alpha_-}{2}~~~~~
\alpha_2 = \frac{\alpha_+ - \alpha_-}{2},
\end{equation}
\begin{equation}
\alpha_+ = 1 - \frac{z_+}{z'_+}~~~~~~~\alpha_-
= \frac{z_+ z'_- - z_- z'_+}{z'_+},
\end{equation}
and
\begin{equation}
\widetilde{O}^{ij}(z_+,z''_+) =
\left( \begin{array}{rr} \delta(z_+ - z''_+)&
  \partial_{z_+}  \delta(z_+ - z''_+)\\
- \theta(z_+ - z''_+)  &\delta(z_+ - z''_+) \end{array}
\right )~.
\end{equation}
\section{Special Cases}
The evolution kernels given above cover a series of limiting cases
which were studied before. These are characterized by
special kinematic conditions
for the matrix elements, as in the case of forward scattering
$\langle p_2| \rightarrow \langle p_1| \equiv \langle p|$ or
the transition from the vacuum state $\langle 0|$ to a hadron state
$\langle p|$.
These conditions lead to a
constraint between the variables $z_+$ and $z_-$, which is defined
by a variable $t = \chi(z_+,z_-)$ and implies related evolution
equations for partition functions depending on the variable $t$.
These partition functions are related to the expectation values of the
operators $Q^q$ and $O^G$ by
\begin{eqnarray}
\left.
\frac{
\langle p_1|O^{q}(-\kappa_- \xx,
\kappa_- \xx)|p_2\rangle}
{(i\xx p_+)}
\right|_{\xx p_- = \tau \xx p_+}
&=&
\int_{-\infty}^{+\infty} dt
e^{-i\kappa_- \xx p_+ t} F_{q}(t)\\
\left.
\frac{
\langle p_1|O^{G}(-\kappa_- \xx,
\kappa_- \xx)|p_2\rangle}
{(i\xx p_+)^2}
\right|_{\xx p_- = \tau \xx p_+}
&=&
\int_{-\infty}^{+\infty} dt
e^{-i\kappa_- \xx p_+ t}~t F_{G}(t)~.
\end{eqnarray}
Here
$\tau =\xx p_- / \xx p_+$
is a general parameter which distinguishes the cases discussed
below.

The evolution equations are~:
\begin{eqnarray}
\label{evo3t}
\mu^2 \frac{d}{d \mu^2} F^{\rm NS}(t) &=&
\frac{\alpha_s(\mu^2)}{2 \pi}
\int_{-\infty}^{+\infty} d t'
V_{ext}^{\rm NS}(t,t',\tau) F^{\rm NS}(t') \\
\label{evo3ta}
\mu^2 \frac{d}{d \mu^2}
\left ( \begin{array}{c}
F^{q}(t) \\
F^{G}(t) \end{array} \right )
 &=&
\frac{\alpha_s(\mu^2)}{2 \pi}
\int_{-\infty}^{+\infty} d t'
\Vvec_{ext}(t,t',\tau)
\left ( \begin{array}{c}
F^{q}(t') \\
F^{G}(t') \end{array} \right )~,
\end{eqnarray}
with $F^{\rm NS} = F^{q_i} - F^{\overline{q}_j}$ and
$V^{\rm NS} = V^{qq}$.
The corresponding
extended kernel $\Vvec(t,t',\tau) = (V^{i,j}(t,t',\tau))$
reads
\begin{eqnarray}
\label{ker3t}
V_{ext}^{ij}(t,t',\tau) &=&
\int_0^1 d \AAA \int_0^{1-\AAA} d \AB
K^{ij}(\AAA,\AB)
\frac{1}{2 \pi} \int_{- \infty}^{+ \infty} d(p_+\xx \kappa_-)
\nonumber\\ & & \times
\left[i p_+\xx (\kappa_- - i \varepsilon)\right]^{a_{ij}}
\frac{{t'}^{a_j}}{t^{a_i}}
\exp\left
\{i p_+ \kappa_- \xx\left[t-(1-\AAA t'-\AB)+ \tau (\AAA -\AB)\right]
\right\},
\end{eqnarray}
with $a_{ij} = a_j - a_i, a_G = 1, a_q =0$.
It obeys the scaling relation
\begin{eqnarray}
V^{ij}_{ext}(t,t',\tau) = \frac{1}{\tau} V_{ext}^{ij}\left(\frac{t}{\tau},
\frac{t'}{\tau},1 \right)~.
\end{eqnarray}
For convenience we write the
general expressions for the evolution kernels in the variables
\begin{eqnarray}
\label{xy}
x= \frac{1}{2}\left ( 1 + \frac{t}{\tau}\right),~~~~~~~~~
y= \frac{1}{2}\left ( 1 + \frac{t'}{\tau}\right)~.
\end{eqnarray}
They are given by~:
\begin{eqnarray}
\label{kerGE1}
V^{qq}_{ext}(t,t',\tau) &=&
\frac{1}{2}
\left \{  V^{qq}(x,y)\rho(x,y) + V^{qq}(\bx,\by) \rho(\bx,\by)
\right \}
\frac{1}{\tau}
\\
V^{qG}_{ext}(t,t',\tau) &=&\frac{1}{2} \left(\frac{2y - 1}{2}
\right)
\left \{  V^{qG}(x,y)\rho(x,y) - V^{qG}(\bx,\by) \rho(\bx,\by)
\right \} \frac{1}{\tau}
\\
V^{Gq}_{ext}(t,t',\tau) &=&\frac{1}{2} \left(
\frac{2}{2x - 1} \right)
\left \{  V^{Gq}(x,y)\rho(x,y)
- \overline{V}^{Gq}(\bx,\by) \rho(\bx,\by)
\right \} \frac{1}{\tau}
\\
\label{kerGE4}
V^{GG}_{ext}(t,t',\tau) &=& \frac{1}{2} \left(
\frac{2y - 1}{2x - 1}  \right)
\left \{
V^{GG}(x,y)\rho(x,y) + V^{GG}(\bx,\by) \rho(\bx,\by) \right \}
\frac{1}{\tau}
\\
\label{kerDGE1}
\Delta V^{qq}_{ext}(t,t',\tau) &=&  V^{qq}_{ext}(t,t',\tau)\\
\Delta V^{qG}_{ext}(t,t',\tau) &=&\frac{1}{2} \left(\frac{2y - 1}{2}
\right)
\left \{\Delta  V^{qG}(x,y)\rho(x,y) - \Delta V^{qG}(\bx,\by)
\rho(\bx,\by)
\right \} \frac{1}{\tau}
\\
\Delta V^{Gq}_{ext}(t,t',\tau) &=&\frac{1}{2} \left (
\frac{2}{2x - 1} \right )
\left \{\Delta  V^{Gq}(x,y)\rho(x,y)
- \Delta \overline{V}^{Gq}(\bx,\by)
\rho(\bx,\by)
\right \} \frac{1}{\tau}
\\
\label{kerDGE4}
\Delta
V^{GG}_{ext}(t,t',\tau) &=& \frac{1}{2} \left(
\frac{2y - 1}{2x - 1}  \right)
\left \{
\Delta V^{GG}(x,y)\rho(x,y) + \Delta V^{GG}(\bx,\by)
\rho(\bx,\by) \right \}
\frac{1}{\tau}
\end{eqnarray}
with
\begin{equation}
\rho(x,y) =
\theta\left(1-\frac{x}{y}\right)
\theta\left(\frac{x}{y}\right)~{\rm sign}(y)~,
\end{equation}
and
\begin{eqnarray}
V^{qq}(x,y) &=& C_F
\left [ \frac{x}{y} - \frac{1}{y} - \frac{1}{(y-x)_+}
+ \frac{3}{2} \delta(x-y) \right ]\\
V^{qG}(x,y) &=& - 2 N_f T_R \frac{x}{y} \left [ 4(1 - x) +
\frac{1 - 2x}{y} \right ] \\
V^{Gq}(x,y) &=&  C_F \left [  \frac{y^2-x^2}{y} + 1 \right ]
~~~~~~~
\overline{V}^{Gq}(x,y) =  C_F \left [  \frac{y^2-x^2}{y} - 1 \right ] \\
V^{GG}(x,y) &=&  C_A
  \left[2\frac{x^2}{y}\left(3-2x + \frac{1-x}{y}\right)
+\frac{1}{(y-x)_+}
   - \frac{y-x}{y^2}\right]
+ \frac{\beta_0}{2} \delta(x-y)\\
\Delta V^{qq}(x,y) &=& V^{qq}(x,y)\\
\Delta V^{qG}(x,y) &=& - 2 N_f T_R \frac{x}{y^2} \\
\Delta V^{Gq}(x,y) &=& C_F \left [1 - \frac{y^2-x^2}{y} \right ]
~~~~~~~
\Delta \overline{V}^{Gq}(x,y) = C_F \left [-1 - \frac{y^2-x^2}{y} \right ]
 \\
\Delta V^{GG}(x,y) &=& C_A
 \left[2\frac{x^2}{y^2}  +\frac{1}{(y-x)_+}
   - \frac{y-x}{y^2}\right]  + \frac{\beta_0}{2} \delta(x - y)
\end{eqnarray}

Note that the kernels given in
eqs.~(\ref{kerGE1}--\ref{kerDGE4}) apply to the {\it full} range of
variables.
The function $V^{qq}_{ext}(t,t',\tau)$ was already derived in
refs.~\cite{ROBA,LEIP}.
For the unpolarized case the kernels (\ref{kerGE1}--\ref{kerGE4}) were
also
calculated in ref.~\cite{RAD} recently, using a different notation.
There separate expressions for different ranges of variables are given.
For $\tau = 1$, corresponding to $\zeta = 1$ in \cite{RAD},
the kernels of \cite{RAD}
for $ 0 \leq x,y \leq 1$
agree with
the result obtained above
up to obvious misprints.
Let us specify now the parameter $\tau$ to derive a series of different
evolution kernels discussed in the literature.
\subsection{\mbox{\boldmath $t' = 1$}}
In this case, which  has been dealt with in ref.~\cite{XJ} recently,
the evolution kernels depend on one variable unlike
the case discussed in the previous section. Furthermore the parameter
$\tau$ introduced above equals to
$- \xi/2$ in the notation of
ref.~\cite{XJ}, and the value of the final state virtuality is
chosen as $q_2^2 = 0$.
Whereas we obtain the corresponding evolution kernels as a special limit
from the general two--variable kernels eqs.~(\ref{eqK1}-\ref{eqK4},
\ref{eqDK1}-\ref{eqDK4}), these kernels were evaluated
in \cite{XJ}
only in a restricted range of variables.
We agree with the results given in \cite{XJ} for
$t > \xi/2$, eqs.~(22, 30).\footnote{A deviating result was reported in
ref.~\cite{PEN} recently studying the unpolarized case. However, as the
kinematic quantities used are not specified in detail, a comparison
is not possible. Note also typographical errors in the color factors.}
These equations, generalizing them
even for $t' \neq 1$, read\footnote{In
ref.~\cite{XJ} the variable $t$ was denoted
by $x$, having a different meaning in our notation.
Therefore we use $t$ instead in eqs.~(\ref{JI1}--\ref{JI4}).}~:
\begin{eqnarray}
\label{JI1}
K^{qq}(t,t',\xi) &=&
C_F \frac{t^2 + {t'}^2 - \xi^2/2}{({t'}^2 - \xi^2/4)(t' - t)_+}
+ \frac{3}{2} \delta(t' - t)
\\
K^{qG}(t,t',\xi) &=&
 T_R N_f \frac{t^2 + (t' - t)^2 - \xi^2/4}{({t'}^2 - \xi^2/4)^2}t'
\\
K^{Gq}(t,t',\xi) &=&
 C_F \frac{{t'}^2 + (t' - t)^2 - \xi^2/4}{t({t'}^2 - \xi^2/4)}\\
K^{GG}(t,t',\xi) &=&  2 C_A \left(\frac{t'}{t}\right)
\frac{1}{({t'}^2 -\xi^2/4)^2}  \left [\frac{({t'}^2 -\xi^2/4)^2}
{(t'-t)_+}       + t'({t'}^2 + \xi^2/4) \right. \nonumber \\
      & &\left. ~~~~~- t(3{t'}^2 -\xi^2/4) - (t'+t)(t'-t)^2 \right]
      + \frac{\beta_0}{2} \delta(t' - t),\\
\Delta K^{qq}(t,t',\xi) &=&  K^{qq}(t,t',\xi) \\
\Delta K^{qG}(t,t',\xi) &=&
T_R N_f \frac{t^2 - (t' - t)^2 - \xi^2/4}{({t'}^2 - \xi^2/4)^2}t' \\
\Delta K^{Gq}(x,\xi) &=&
C_F \frac{t' - (t' - t)^2 - \xi^2/4}{t({t'}^2 - \xi^2/4)} \\
\label{JI4}
\Delta K^{GG}(x,\xi) &=& 2 C_A \left(\frac{t'}{t} \right)
\frac{1}{({t'}^2 -\xi^2/4)^2} \left [\frac{({t'}^2 -\xi^2/4)^2}
{(t'-t)_+}  +       t'({t'}^2 + \xi^2/4) \right.
\nonumber \\       & & \left.
 - t(3{t'}^2 -\xi^2/4) -2t'(t'-t)^2 \right ]
      + \frac{\beta_0}{2} \delta(t' - t)~.
\end{eqnarray}

\subsection{The Brodsky--Lepage Limit}
For $\tau = \pm 1$ the equations~(\ref{kerGE1}--\ref{kerDGE4})
transform into the limit $\langle p_2| \rightarrow \langle p|,
\langle p_1| \rightarrow \langle 0|$, which is known as the
Brodsky--Lepage~\cite{BL} and Efremov--Radyushkin~\cite{ER}
case.\footnote{Several independent calculations of the evolution
kernels for the meson wave functions were performed in refs.~\cite{BLX}.}
This limit may be performed {\it formally} leaving $p_1 \rightarrow 0$,
leading to correct results, cf.~\cite{ROBA}.
The corresponding evolution equations are obtained using as
variables $x$ and $y$, eq.~(\ref{xy}), in (\ref{evo3t},\ref{evo3ta}).
\subsection{The Altarelli--Parisi Limit}
We consider the case of forward scattering $p_2 = p_1 \equiv p$. 
The corresponding
evolution kernels can be obtained after some calculation from
eqs.~(\ref{kerGE1}--\ref{kerDGE4})
in the limit $\tau \rightarrow 0$ or, alternatively, by a direct
calculation in the forward scattering case using the kernels
eqs.~(\ref{eqK1}--\ref{eqDK4}), which is performed in the following.

The quark and
gluon distributions are related to the operator expectation values by
\begin{eqnarray}
f^q(z,\mu) &=& \frac{1}{2 \pi} \int_{- \infty}^{+ \infty} d(2p\xx \kappa_-)
\langle p|O^q|p\rangle(\kappa_-,\mu)
\frac{e^{2i p\xx \kappa_-}}{2ip\xx}
\\
zf^G(z,\mu) &=& \frac{1}{2 \pi}
\int_{- \infty}^{+ \infty} d(2p\xx \kappa_-)
\langle p|O^G|p\rangle(\kappa_-,\mu)
\frac{e^{2i p\xx \kappa_-}}{(2ip\xx)^2}~.
\end{eqnarray}
The respective polarized parton densities are obtained by replacing
$f^{q,G}(z)$ by $\Delta f^{q,G}(z)$ and $O^{q,G}$ by $O^{q,G}_5$.
The evolution equation for the operators  $O^{q,G}_{(5)}$,
eqs.~(\ref{evo1},\ref{evo2}), can be rewritten in this limit by
\begin{eqnarray}
\label{evoAP}
\mu^2 \frac{d}{d \mu^2} f^{\rm NS}(z, \mu) &=& \frac{\alpha_s(\mu^2)}
{2 \pi} \int_{- \infty}^{+ \infty} \frac{d z'}{|z'|}
\widehat{P}^{\rm NS}\left (
\frac{z}{z'} \right) f^{\rm NS}(z, \mu),  \\
\mu^2 \frac{d}{d \mu^2} \left (\begin{array}{c}
f^q(z, \mu) \\
f^G(z,\mu) \end{array} \right ) &=&
\frac{\alpha_s(\mu^2)}{2 \pi}
\int_{-\infty}^{+\infty}
\frac{d z'}{|z'|}
\widehat{\Pvec} \left (\frac{z}{z'} \right)
\left (\begin{array}{c}
f^q(z, \mu) \\
f^G(z,\mu) \end{array} \right )~.
\end{eqnarray}
The splitting functions $\widehat{P}^{ij}(z)
(\Delta \widehat{P}^{ij}(z))$
are related to the kernels $K^{ij}(\alpha_1, \alpha_2) (\Delta
K^{ij}(\alpha_1, \alpha_2))$ by
\begin{eqnarray}
\label{APSP}
\widehat{P}^{ij}(z) &=& P^{ij}(z) \theta(z) \theta(1 - z)
\nonumber\\
P^{ij}(z) &=& \int_{- \infty}^{+ \infty}
du \widehat{O}^{ij}(u,z) \int_0^1 d\xi (1 - u)
\widehat{K}^{ij}(\alpha_1, \alpha_2) \theta(1 - u) \theta(u),
\end{eqnarray}
with
$\alpha_1  = \xi (1 - u)$, $\alpha_2 =  (1 - \xi)(1 - u)$, and
\begin{equation}
\widehat{\Kvec}    = \left( \begin{array}{rr} K^{qq}
&(1/   \kappa_-) K^{qG}   \\
(\kappa_- -i \varepsilon)  K^{Gq}    & K^{GG}    \end{array}
\right),~~~~~~~\widehat{O}^{ij}(u,z) =
\left( \begin{array}{rr} \delta(z - u)~~~&
\partial_z  \delta(z - u)~~~\\
- \theta(z - u)/z  &\delta(z-u)/z \end{array}
\right )~.
\end{equation}
$P^{ij}(z) (\Delta P^{ij}(z))$ are the well--known
Altarelli--Parisi splitting functions~\cite{AP} for unpolarized and
polarized deep--inelastic scattering
\begin{eqnarray}
P^{qq}(z) &=&
C_F \left (\frac{1 + z^2}{1 - z} \right)_+
\\
P^{qG}(z) &=& 2 N_f T_R \left [z^2 + (1 - z)^2 \right]\\
P^{Gq}(z) &=& C_F \frac{1 + (1 - z)^2}{z}\\
P^{GG}(z) &=& 2 C_A \left [ \frac{1}{z} + \frac{1}{(1 - z)}_+ -2
+ z(1-z)  \right ] + \frac{\beta_0}{2} \delta(1 - z),\\
\Delta P^{qq}(z) &=& P^{qq}(z)\\
\Delta P^{qG}(z) &=& 2 N_f T_R \left [z^2 - (1 - z)^2 \right ] \\
\Delta P^{Gq}(z) &=& C_F \frac{1 - (1 - z)^2}{z}\\
\Delta P^{GG}(z) &=& 2 C_A \left [ 1 - 2 z + \frac{1}{(1 - z)}_+ \right ]
+ \frac{\beta_0}{2} \delta(1 - z).
\end{eqnarray}
In deriving eq.~(\ref{APSP}) it is useful to apply the relations
\begin{equation}
\theta(x) = \lim_{\varepsilon \rightarrow 0^+}
\frac{1}{2\pi}
\int_{- \infty}^{+ \infty} d\xi \frac{e^{ix\xi}}{i\xi + \varepsilon},
~~~~~\delta^{(k)}(x) =
\frac{1}{2\pi}
\int_{- \infty}^{+ \infty} d\xi~(i\xi)^{k} e^{ix\xi},
\end{equation}
which are valid for tempered distributions~\cite{VLA}.

\section{Conclusions}
The evolution kernels for the twist~2 light--ray operators both for
the case of unpolarized and polarized deep inelastic non--forward
scattering were derived for the flavor non--singlet and singlet cases.
In general  the partition  functions depend on two distribution
variables.
One may study as well specialized evolution equations in one distribution
variable implying external
 constraints, covering the case of evolution
equations for  non--forward parton densities. In this way,
among various others,
also the well--known evolution equations  as the Brodsky--Lepage or
Altarelli--Parisi equations can be obtained.

\vspace{3mm}
\noindent
{\bf Acknowledgement}~We would like to thank Paul S\"oding for his
constant support of the  project and D. M\"uller for discussions on
the present topic.

\end{document}